\documentclass[a4paper,11pt]{article}
\usepackage{pos}

\newcommand{\be}{\begin{eqnarray}}
\newcommand{\ee}{\end{eqnarray}}

\title{Physics and Astrophysics of Black Holes with eXTP}
\ShortTitle{Physics and Astrophysics of Black Holes with eXTP}

\author*[a,b]{Cosimo~Bambi}

\affiliation[a]{Center for Astronomy and Astrophysics, Department of Physics, Fudan University,\\
2005 Songhu Road, Shanghai 200438, China}

\affiliation[b]{School of Natural Sciences and Humanities, New Uzbekistan University,\\
Movarounnahr Street 1, Tashkent 100000, Uzbekistan}

\emailAdd{bambi@fudan.edu.cn}

\abstract{The enhanced X-ray Timing and Polarimetry (eXTP) mission will combine spectral, timing, and polarimetric techniques to study accreting black holes, measure their masses and spins, and test Einstein's theory of General Relativity in the strong-field regime. In this contribution, I review the capabilities of eXTP to advance our current understanding of black hole physics and astrophysics.}

\FullConference{Mondello Workshop 2026 Frontier Research in Astrophysics – V (FRAPWS2026)\\
15-20 June 2026\\
Palermo, Italy\\}

\begin{document}

\maketitle


\section{Introduction}\label{s-intro}

The enhanced X-ray Timing and Polarimetry (eXTP) mission is a Sino-European scientific space project led by the Institute of High Energy Physics of the Chinese Academy of Sciences~\cite{Zhang:2025iae}. Designed to study neutron stars and black holes through X-ray spectral, timing, and polarimetric observations, the mission is currently scheduled for launch at the beginning of 2030.

In its new baseline configuration, the scientific payload of eXTP comprises three main instruments: the Spectroscopic Focusing Array (SFA), the Polarimetry Focusing Array (PFA), and the Wide-band and Wide-field Camera (W2C)~\cite{Zhang:2025iae}.
\begin{enumerate}
\item The SFA consists of five SFA-T (Spectroscopic Focusing Array-Timing) telescopes and one SFA-I (Spectroscopic Focusing Array-Imaging) telescope, all covering the energy range 0.5-10~keV and with a designed angular resolution of $\sim$1' (half-power diameter, HPD). The SFA-T telescopes are equipped with silicon-drift detectors (SDDs): their total effective area is 2750~cm$^2$ at 1.5~keV (1670~cm$^2$ at 6~keV), the spectral resolution is $\sim$180~eV at 6~keV, and the time resolution is 10~$\mu$s. The SFA-I telescope is equipped with pn-CCD detectors; it provides an effective area of 550~cm$^2$ at 1.5~keV (330~cm$^2$ at 6~keV), a spectral resolution of $\sim$180~eV at 1.5~keV, and a field of view 18'$\times$18'.
\item The PFA consists of three identical telescopes optimized for X-ray imaging polarimetry, covering the energy range 2-8~keV. The angular resolution is $\sim$30'' (HPD) over a field of view of 9.8'$\times$9.8'. The total effective area is 250~cm$^2$ at 3~keV, and the expected energy resolution is $\sim$20\% at 6~keV. The instrument achieves an expected minimum detectable polarization at the 99\% confidence level of about 2\% in 1~Ms for a milli-Crab source. 
\item The W2C is characterized by a field of view of 1500~deg$^2$ (Full-Width Zero Response, FWZR), an energy range of 30-600~keV, an angular resolution of 20', and an energy resolution of $\sim$30\% at 60~keV. 
\end{enumerate}
Further details on the eXTP instruments can be found in Ref.~\cite{Zhang:2025iae}.

Last year, the eXTP Collaboration published six white papers: a comprehensive overview of the mission and its instruments~\cite{Zhang:2025iae}; the science of matter at very high densities in neutron stars~\cite{Li:2025uaw}; the science of strong gravity around black holes~\cite{Bu:2025jxy}; the science of strong magnetic fields in neutron stars~\cite{Ge:2025upe}; the contribution of eXTP to time-domain and multi-messenger astrophysics~\cite{Yi:2025fpm}; and the observatory science enabled by eXTP~\cite{Zhou:2025pth}. This contribution summarizes the third white paper, which focuses on the ability of eXTP to probe strong gravity with black holes.


\section{Disk-Corona Model}\label{s-pop}

\begin{figure}[t]
\centering
\includegraphics[width=0.45\linewidth]{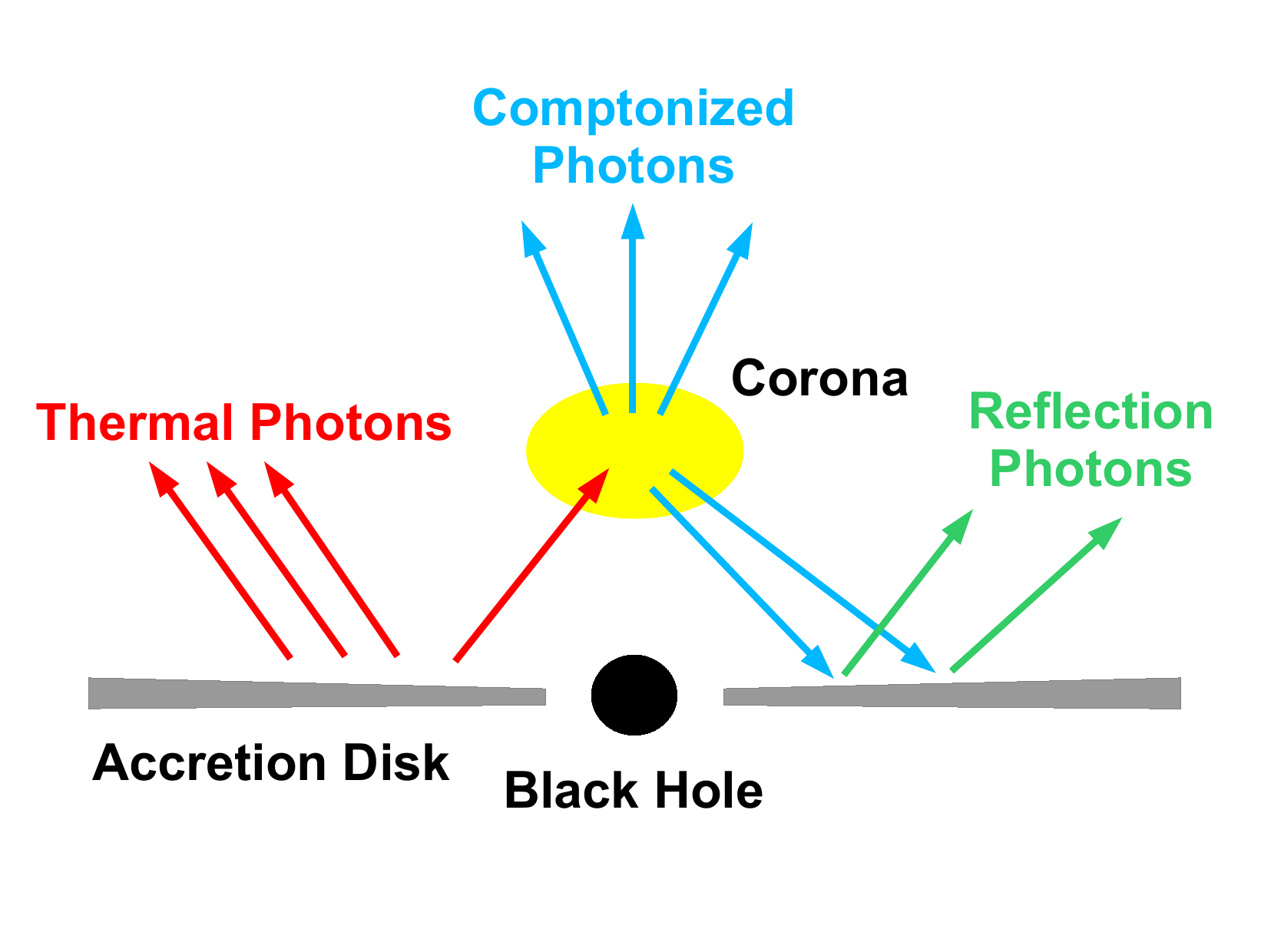}
\hspace{0.8cm}
\includegraphics[width=0.45\linewidth]{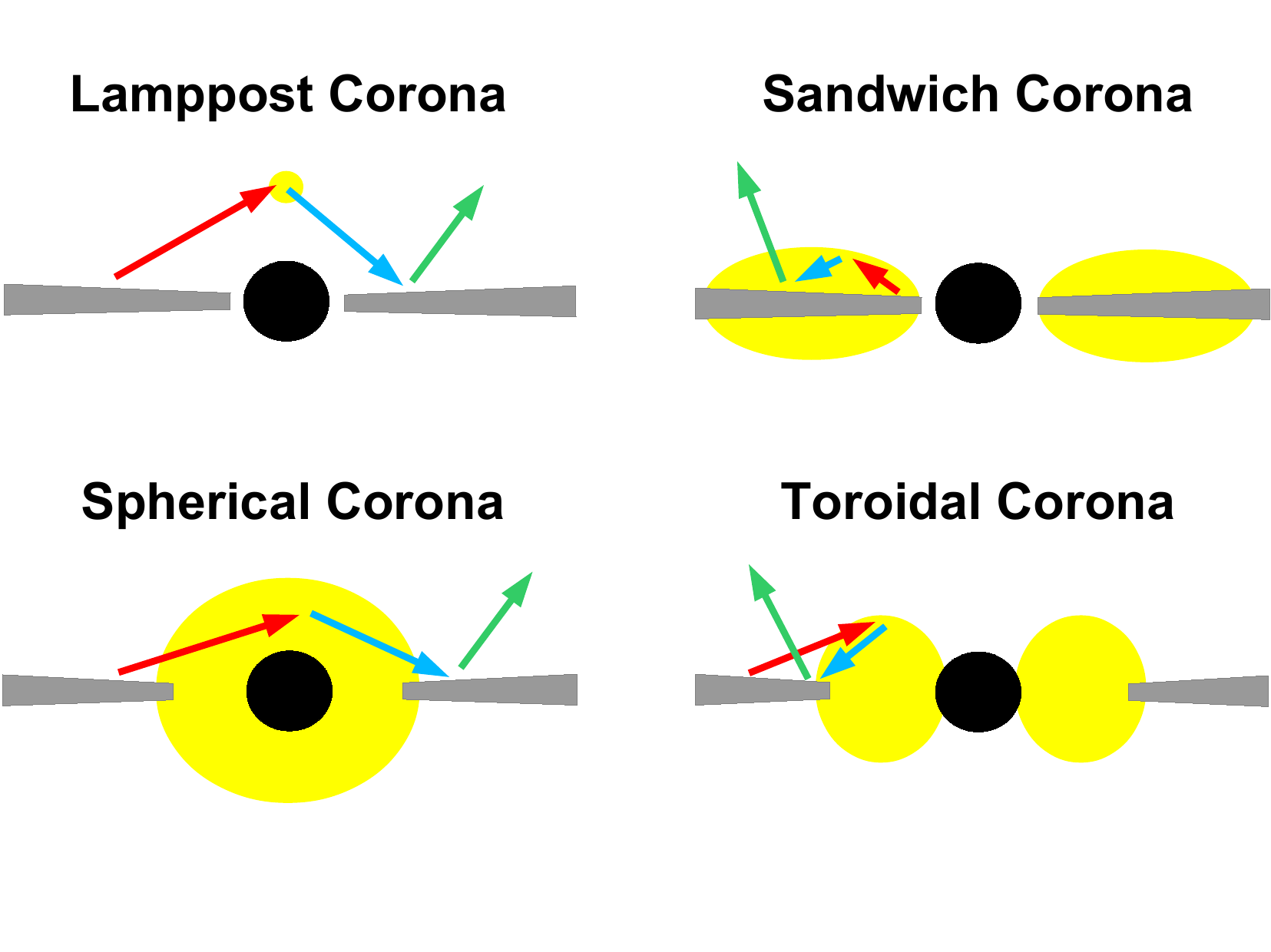}
\vspace{-0.3cm}
\caption{Sketch of the disk-corona model (left) and examples of possible coronal geometries (right). Figure from Ref.~\cite{Bambi:2024hhi}.}
\label{f-c}
\end{figure}

The analysis of X-ray radiation emitted from the strong-gravity region around black holes provides a powerful tool for studying the physics and astrophysics of these objects. The prototypical astrophysical system is illustrated in Fig.~\ref{f-c} and is commonly referred to as the disk-corona model~\cite{refc1,refc2}. The black hole may be either a stellar-mass black hole in an X-ray binary or a supermassive black hole in an active galactic nucleus. The essential feature is that the black hole is surrounded by a Keplerian, geometrically thin, and optically thick accretion disk. This disk is ``cold'' because it efficiently radiates: its thermal spectrum scales as $M^{-1/4}$, where $M$ is the black hole mass. Consequently, the spectrum peaks in the soft X-ray band (0.1-10~keV) for stellar-mass black holes in X-ray binaries, and in the UV/optical band (1-100~eV) for supermassive black holes in active galactic nuclei~\cite{Page:1974he,Shakura:1972te}. The {\it corona} is a ``hot'' ($\sim$100~keV) plasma located near the black hole and the inner region of the accretion disk; however, its exact morphology remains poorly understood. Fig.~\ref{f-c} depicts several possible coronal geometries, including the lamppost corona (a compact corona along the black hole spin axis), the sandwich corona (a thin layer above the disk), a spherical corona, and a toroidal corona; further details can be found in, e.g., Ref.~\cite{Bambi:2024hhi} and references therein.

Because the disk is cold and the corona is hot, thermal photons from the disk can inverse-Compton scatter off free electrons in the corona. The spectrum of these Comptonized photons is typically well approximated by a power law with a high-energy and a low-energy cutoff. A fraction of the Comptonized photons illuminates the disk, where they undergo Compton scattering and absorption followed by fluorescent emission. The result is a reflection spectrum.

The {\it non-relativistic reflection spectrum} -- that is, the reflection spectrum in the rest frame of the disk material -- is characterized by narrow fluorescent emission lines in the soft X-ray band and a Compton hump peaking at 20-50~keV~\cite{Ross:2005dm,Garcia:2010iz}. The {\it relativistic reflection spectrum}, conversely, is the reflection spectrum of the entire disk as observed far from the source. It is obtained by summing the spectra emitted from each point of the disk, after accounting for the propagation of photons from the disk through the strong gravitational field of the black hole to the distant observer in the flat faraway region~\cite{Bambi:2017khi}. These photons are therefore affected by Doppler boosting (due to the Keplerian motion of the disk material) and gravitational redshift (because the photons must climb out of the black hole's gravitational potential).

eXTP will be capable of studying both stellar-mass black holes in X-ray binaries and supermassive black holes in active galactic nuclei. It is expected to address the following open questions:
\begin{enumerate}
\item Are astrophysical black holes the Kerr black holes predicted by Einstein's theory of General Relativity?
\item What is the spin distribution of stellar-mass black holes in X-ray binaries? Is it consistent with that observed in binary black holes via gravitational waves, or do black hole X-ray binaries and binary black holes belong to two distinct populations?
\item What is the spin distribution of supermassive black holes in active galactic nuclei, and what mechanisms are responsible for it?
\item What is the geometry of the corona? How does it form, and how does it evolve over time?
\item What mechanism is responsible for quasi-periodic oscillations (QPOs) -- narrow features at characteristic frequencies commonly observed in the X-ray power density spectra of black holes and neutron stars?
\item What mechanism is responsible for the powerful jets commonly observed in both stellar-mass black holes in X-ray binaries and supermassive black holes in active galactic nuclei?
\end{enumerate}
By combining spectral, timing, and polarimetric techniques, eXTP promises to significantly improve our understanding of black holes.


\section{Measuring Black Hole masses}\label{s-m} 

eXTP will be able to measure black hole masses using at least three methods: $i)$ by combining the analysis of the thermal and reflection spectra for the same source (applicable to stellar-mass black holes); $ii)$ through X-ray reverberation mapping (for supermassive black holes); and $iii)$ via measurements of X-ray variability and QPO frequencies (for both stellar-mass and supermassive black holes).

Within the Novikov-Thorne model~\cite{Page:1974he,NT72}, the thermal spectrum of the disk is determined by five parameters: the black hole mass, the black hole spin, the mass accretion rate, the distance to the source, and the inclination angle of the disk. If independent measurements of the black hole mass, distance, and disk inclination are available, the thermal spectrum can be fitted to infer the black hole spin and the mass accretion rate. This is the so-called continuum-fitting method~\cite{Zhang:1997dy,McClintock:2013vwa}, which has been used to measure the spins of approximately 20~stellar-mass black holes in X-ray binaries~\cite{Draghis:2022ngm}. X-ray reflection spectroscopy, on the other hand, refers to the analysis of the reflection spectrum and constitutes another technique for measuring black hole spins~\cite{Fabian:1989ej,Bambi:2020jpe}. The reflection spectrum is independent of the black hole mass, the mass accretion rate, and the source distance; the fit can therefore directly infer the black hole spin and the disk inclination angle. When the spectrum of a source simultaneously exhibits a prominent thermal component and strong reflection features (see, for instance, Refs.~\cite{Parker:2016ltr,Riaz:2023yde}), the continuum-fitting method and X-ray reflection spectroscopy can be combined to measure the black hole mass. In this approach, the reflection spectrum provides the spin and inclination, which are then used as inputs to the continuum-fitting method to determine the mass.

In X-ray reverberation mapping, we exploit the geometry of the lamppost corona model. When the corona produces a flare, a time lag can be observed between the Comptonized photons -- which travel directly from the corona to the observer -- and the reflected photons, which are produced when the Comptonized photons illuminate the disk and then propagate from the disk to the observer. Simulations reported in Ref.~\cite{Bu:2025jxy} show that eXTP/SFA achieves a significantly higher signal-to-noise ratio than XMM-Newton/EPIC-pn at the iron line, enabling mass measurements of supermassive black holes in active galactic nuclei.

The short-term X-ray variability of active galactic nuclei is known to be anticorrelated with the masses of their supermassive black holes~\cite{Paolillo:2025khv}. eXTP/SFA will benefit from its long uninterrupted exposures, large effective area, and high sensitivity. In particular, it will be able to measure the masses of active galactic nuclei that are too faint for current X-ray missions~\cite{Bu:2025jxy}.

QPO frequencies could potentially provide precise measurements of the masses of stellar-mass black holes in X-ray binaries. In most theoretical models, QPO frequencies are associated in some way with the fundamental frequencies of test particles in the Kerr background~\cite{Bambi:2017khi}. At present, however, the correct mechanism -- if any -- is unknown, and different models relate QPO frequencies to the fundamental frequencies in different ways, leading to discrepant results. Nevertheless, if the correct model were identified, measurements of QPO frequencies could yield very precise determinations of both the black hole mass and spin.


\section{Measuring Black Hole Spins}\label{s-s}

eXTP will be able to measure black hole spins using at least four methods: $i)$ X-ray reflection spectroscopy, $ii)$ the continuum-fitting method, $iii)$ X-ray polarimetry, and $iv)$ QPO frequencies.

\begin{figure}[t]
\centering
\includegraphics[width=0.75\linewidth]{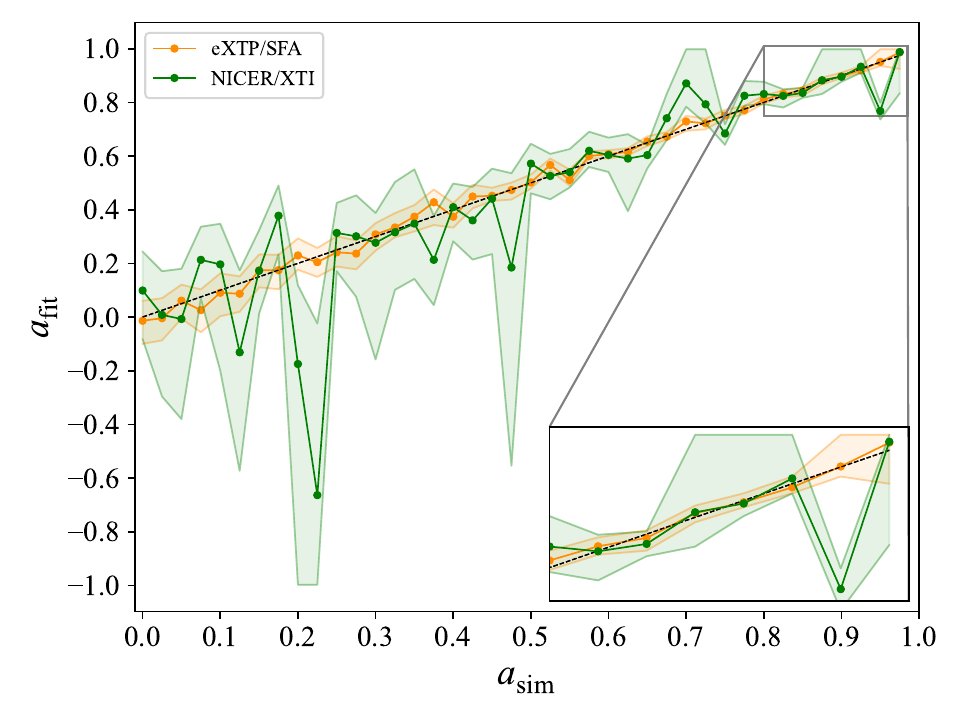}
\vspace{-0.3cm}
\caption{Ability of eXTP/SFA and NICER/XTI to recover black hole spins. The simulations assume bright black hole X-ray binaries (flux $2 \cdot 10^{-8}$~erg~cm$^{-2}$~s$^{-1}$ in the 2-10~keV band) and an exposure time of 30~ks. We employ the {\tt relxill} model~\cite{Dauser:2013xv,Garcia:2013lxa} with the following input parameters: photon index $\Gamma = 1.7$, cutoff energy $E_{\rm cut} = 300$~keV, inner disk radius at the innermost stable circular orbit $R_{\rm in} = R_{\rm ISCO}$, disk inclination angle $i = 45^\circ$, emissivity profile described by a power-law with emissivity index $q = 3$, iron abundance $A_{\rm Fe} = 1$ (Solar), ionization parameter $\log\xi = 3$ ($\xi$ in erg~cm~s$^{-1}$), and reflection fraction $R_{\rm f} = 1$. We fit the simulated data in the range 4-10~keV , as current reflection models are unsuitable for fitting the data of black hole X-ray binaries below 4~keV owing to the high temperature of the disk. Figure courtesy of Abdurakhmon Nosirov.}
\label{f-spin}
\end{figure}

X-ray reflection spectroscopy is currently the leading technique for measuring the spins of stellar-mass black holes in X-ray binaries, and the only available technique for measuring the spins of supermassive black holes~\cite{Bambi:2020jpe,Draghis:2022ngm}. eXTP can provide more precise and accurate spin measurements than current X-ray missions, thanks to the larger effective area of the SFA instrument and the possibility of simultaneous X-ray polarization measurements. Fig.~\ref{f-spin} illustrates the capability of eXTP/SFA and NICER/XTI to recover the black hole spin. The simulations assumes bright black hole X-ray binaries (flux $2 \cdot 10^{-8}$~erg~cm$^{-2}$~s$^{-1}$ in the 2-10~keV band) and 30~ks observations. The spin parameter used in the simulations, $a_{\rm sim}$, is reported along the $x$-axis, while the spin inferred from the analysis, $a_{\rm fit}$, is shown on the $y$-axis. The black-dashed line corresponds to $a_{\rm fit} = a_{\rm sim}$. The shaded areas represent the 90\% confidence intervals. Fig.~\ref{f-spin} clearly demostrates that the eXTP/SFA measurements are more precise than those obtained with NICER/XTI. X-ray polarization measurements can further help improve the accuracy of these spin measurements, as they help to better constrain the geometry of the corona and, more generally, of the entire system.

The capability of eXTP/SFA to measure black hole spins with the continuum-fitting method is discussed in Ref.~\cite{Bu:2025jxy} and compared with that of NICER/XTI. As in the case of X-ray reflection spectroscopy, the larger effective area of eXTP/SFA permits more precise spin measurements (the uncertainty is reduced by a factor of $\sim$3 compared to NICER/XTI measurements).

X-ray polarimetry can also measure the spins of stellar-mass black holes in X-ray binaries, provided that the source is in the soft state. The simulations reported in Ref.~\cite{Bu:2025jxy} examine 500~ks observations with eXTP/PFA of bright black hole X-ray binaries and demonstrate the ability of eXTP/PFA to measure black hole spins. Even more precise and accurate spin measurements can be expected from simultaneous observations with eXTP/SFA and eXTP/PFA.

As in the case of black hole mass measurements, QPO frequencies could potentially provide precise measurements of the spins of stellar-mass black holes in X-ray binaries. Unfortunately, the exact mechanism responsible for the observed QPOs in X-ray binaries remains unknown, and consequently all such measurements are model-dependent: they may be very precise, but they may not be accurate. However, thanks to the possibility of combining spectral, timing, and polarimetric analyses of the same source, eXTP will be able to test a number of QPO models and, hopefully, identify the correct one.


\section{Testing General Relativity}\label{s-gr}

In four-dimensional General Relativity, and in the absence of exotic matter fields, the final product of gravitational collapse should be a Kerr black hole. Deviations from the Kerr metric are possible in the presence of exotic matter fields, if General Relativity is not the correct theory of gravity, or as a result of macroscopic quantum gravity effects. Testing whether the spacetime geometry around a black hole is described by the Kerr solution is therefore a crucial test of General Relativity in the strong-field regime~\cite{Bambi:2015kza}. The techniques used to measure black hole spins can also be employed to test the Kerr geometry: X-ray reflection spectroscopy~\cite{Cao:2017kdq,Tripathi:2018lhx,Tripathi:2020yts,Bambi:2022dtw}, the continuum-fitting method~\cite{Zhou:2019fcg,Tripathi:2020qco}, X-ray polarimetry~\cite{Krawczynski:2012ac,Liu:2015ibq}, and QPOs~\cite{Bambi:2012pa,Bambi:2013fea}.

\begin{figure}[t]
\centering
\includegraphics[width=0.75\linewidth]{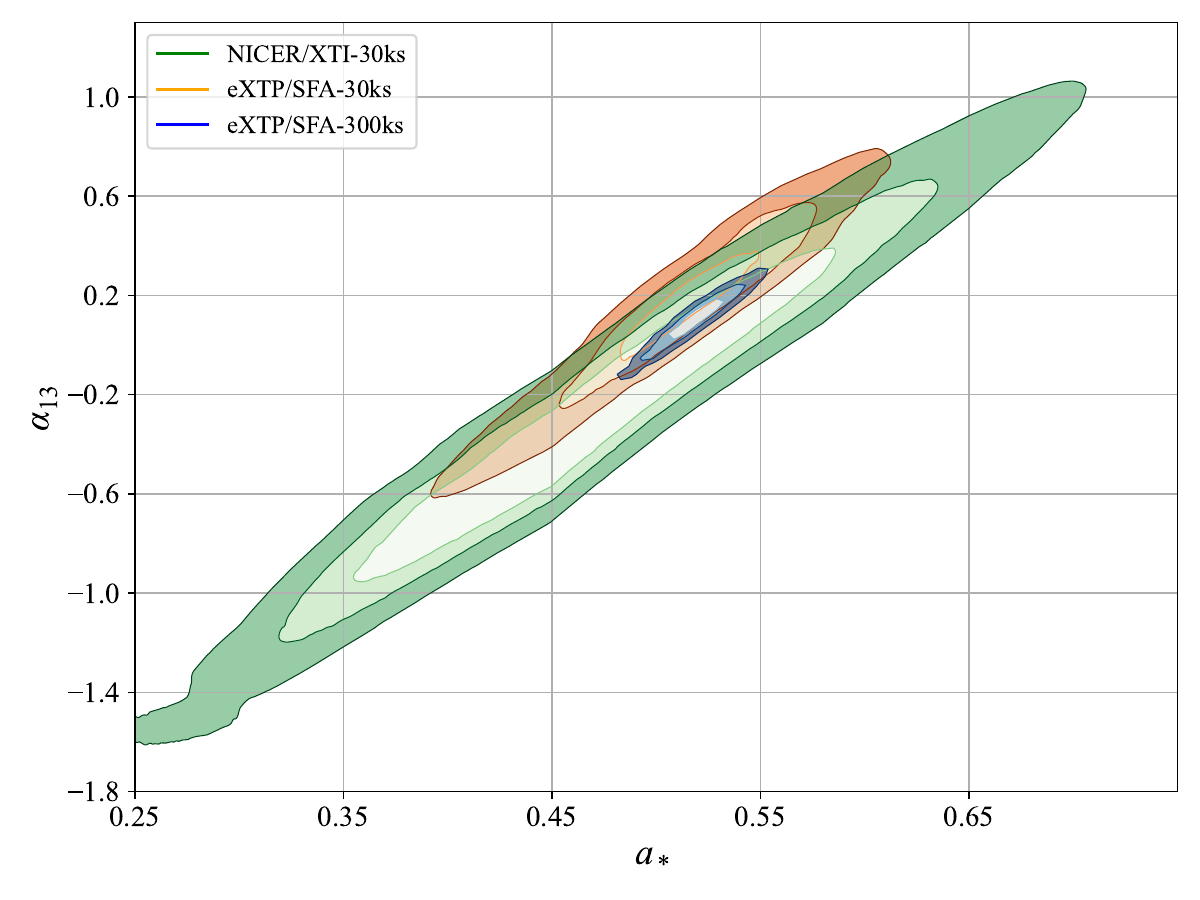}
\vspace{-0.3cm}
\caption{Ability of eXTP/SFA and NICER/XTI to test the Kerr geometry. We show the measurements of the spin parameter $a_*$ and of the Johannsen deformation parameter $\alpha_{13}$ ($\alpha_{13} = 0$ corresponds to the Kerr solution, while $\alpha_{13} \neq 0$ indicates a deviation from the Kerr geometry) for a 30~ks observation with NICER/XTI (in green), a 30~ks observation with eXTP/SFA (in orange), and a 300~ks observation with eXTP/SFA (in blue). For each measurement, we show the 68\%, 90\%, and 99\% confidence level contours for two parameters. In the simulation, we assume a bright black hole X-ray binary (flux $2 \cdot 10^{-8}$~erg~cm$^{-2}$~s$^{-1}$ in the 2-10~keV band) with spin parameter $a_{\rm sim} = 0.5$, vanishing deformation parameter $\alpha_{13} = 0$ (Kerr geometry), and disk inclination angle $i = 70^\circ$. Figure courtesy of Abdurakhmon Nosirov.}
\label{f-gr}
\end{figure}

X-ray reflection spectroscopy is currently the leading technique for testing the Kerr geometry. Fig.~\ref{f-gr} shows the ability of eXTP/SFA and NICER/XTI to constrain deviations from the Kerr metric. For the simulations and the fits, we used the model {\tt relxill\_nk}~\cite{Bambi:2016sac,Abdikamalov:2019yrr,Abdikamalov:2020oci}, which is an extension of the {\tt relxill} package for non-Kerr spacetimes. The simulations assume a Kerr black hole with spin parameter $a_{\rm sim} = 0.5$ and disk inclination angle $i = 70^\circ$. We consider a bright black hole X-ray binary and present the results for a 30~ks observation with NICER/XTI (in green), a 30~ks observation with eXTP/SFA (in orange), and a 300~ks observation with eXTP/SFA (in blue). We fit the data with the Johannsen metric, where the deformation parameter $\alpha_{13}$ quantifies possible deviations from the Kerr solution ($\alpha_{13} = 0$ corresponds to the Kerr metric, while $\alpha_{13} \neq 0$ indicates a non-Kerr geometry). Fig.~\ref{f-gr} shows the constraints on the black hole spin and the deformation parameter $\alpha_{13}$ after marginalizing over all other parameters. eXTP/SFA can provide more stringent constraints on $\alpha_{13}$ than NICER/XTI, owing to its larger effective area. The ability of eXTP/SFA to constrain $\alpha_{13}$ can be further improved with longer exposure times.

The continuum-fitting method, by itself, cannot constrain the spacetime geometry around a black hole because of a degeneracy between the measurement of the black hole spin and possible deviations from the Kerr solution (see, for instance, Ref.~\cite{Tripathi:2020qco}). However, if a source exhibits both a prominent thermal component and strong relativistic reflection features -- or, alternatively, if spectra with a prominent thermal component and spectra with strong relativistic reflection features of the same source are combined -- the continuum-fitting method can assist the reflection-spectrum analysis in providing more stringent constraints (see, for instance, Refs.~\cite{Tripathi:2020dni,Tripathi:2021rqs,Zhang:2021ymo}).

The analysis of the polarization of the spectrum of a source can also be used to test the Kerr geometry around a black hole. While the constraining power is expected to be weak, as in the case of the continuum-fitting method, we can expect to combine polarization analysis with the spectral analysis of the reflection and thermal components to improve upon the constraints that would be obtained from the reflection and thermal components alone.

In the case of QPOs, the same problem arises as in the measurements of black hole masses and spins. If the exact mechanism responsible for QPOs were known, very stringent tests of the Kerr metric could be obtained. At present, however, the exact mechanism remains unknown, and consequently every test of the Kerr metric using QPOs is model-dependent.


\section{Concluding Remarks}\label{s-cr}

eXTP is a Sino-European X-ray mission currently scheduled for launch at the beginning of 2030. One of its primary scientific goals is to study the strong gravitational fields of stellar-mass black holes in X-ray binaries and supermassive black holes in active galactic nuclei. The key feature of this mission is its ability to combine X-ray spectral, timing, and polarimetric techniques. A comprehensive study of how eXTP can advance our current understanding of black holes is reported in Ref.~\cite{Bu:2025jxy}.

However, it is important to emphasize that, in order to fully exploit the capabilities of eXTP, more advanced models for data analysis need to be developed. This is particularly true for X-ray polarimetry, for which only a few simple models are currently available. With approximately 3-4 years remaining before launch, there is a timely opportunity to bridge this gap.


\acknowledgments

This work was supported by the National Natural Science Foundation of China (NSFC), Grant No.~W2531002.


\end{document}